\documentclass[pre, preprint]{revtex4}
\usepackage{graphicx}
\usepackage{amssymb}

\begin{document}
\title{Bouchaud-M\'ezard model on a random network}
\author{Takashi Ichinomiya}
\email{tk1miya@gifu-u.ac.jp}
\affiliation{Department of Biomedical Informatics, Gifu University
Graduate School of Medicine, Yanagido 1-1, Gifu 501-1194, Japan}
\affiliation{PRESTO, Japan Science and Technology Agency, 4-1-8 Honcho
Kawaguchi, Saitama 332-0012, Japan}
\date{\today}                                           
\begin{abstract}
 We studied the Bouchaud-M\'ezard(BM) model, which was introduced to
 explain  Pareto's law in a real  economy, on  a random network.
 Using  "adiabatic and independent" assumptions, we analytically
 obtained  the stationary probability distribution function of wealth.
 The results shows that wealth-condensation, indicated by the
 divergence of the variance of wealth, occurs at a larger $J$ than
 that obtained by the mean-field theory, where $J$ represents the strength
 of interaction between agents. We compared  our results with numerical
 simulation results and found that they were in  good agreement.
\end{abstract}
 \maketitle

\section{Introduction}
 Researchers in the field of complex networks agree that a change in the
 network topology induces  a critical change  in dynamics. 
 Pastor-Satorras and Vespignani first 
 showed  the absence of an epidemic threshold in a scale-free
 network\cite{Pastor-Satorras2001}, following which many researchers
 have focused  on the dynamics on complex networks, such as
 synchronization\cite{Nishikawa2003,Ichinomiya2004}, pattern
 formation\cite{Nakao2008}, and other phenomena.

 In this study, we focus on  the  Bouchaud-M\'ezard(BM) model
 on a complex network\cite{Bouchaud2000}. It is known that the
 wealth distribution in a real economy exhibits a power-law
 behavior, called Pareto's law\cite{Pareto}. With a view to  this power
 law, Bouchaud and M\'ezard proposed a  model, given by the following
 Stratonovich stochastic  differential equation.
\begin{equation}
 dx_i= \frac{J}{N}\sum_{j=1}^N (x_j-x_i) dt  + \sqrt 2 \sigma x_i \circ dW_i \label{BM-original}, 
\end{equation}
where $x_i$, $N$, $J$, and $\sigma^2$ represent the wealth of the $i$-th agent,
number of agents, coupling between agents, and variation of noise,
 respectively. 
 In this model, the evolution of the wealth is determined by two processes:
 exchange of wealth and a random multiplicative process, respectively
 described by the first and second term in the
 right-hand side of Eq.(\ref{BM-original}). 
 Bouchaud and M\'ezard
 analyzed this model using the mean-field theory and  calculated the
 probability density function(PDF) of wealth. They showed  that the
 stationary distribution
 of normalized wealth $x_i/\langle x \rangle$, where $\langle\cdots
 \rangle$ represents the
 average over all agents, exhibits the  power-law behavior. They also
 found that  ``wealth-condensation,''   which is indicated by the
 divergence of the variance of $x_i/\langle x \rangle$, occurs at
 $J \le J_c= \sigma^2$.
 The divergence of the variance  implies that wealth condenses to a
 few rich agents.
On the other hand, if $J> J_c$, the variance remains finite, and  the
 wealth of many agents is close to the average.

 In the original BM model, all agents are coupled with each other. However,
 in a  real economy, agents can exchange their wealth with a  limited
 number of agents. 
 Therefore, it is natural to extend the BM model on a  complex network in
 which the number of neighbors is  limited. Some
 studies have already dealt with  this subject. In their  original study
 on the BM-model\cite{Bouchaud2000},
  Bouchaud and M\'ezard carried out  numerical simulations on a regular
 random graph to estimate the exponent of the power-law behavior. They
 reported that the exponent
 obtained from numerical simulation becomes smaller than that obtained
 from  mean-field theory.
 Some studies have also reported on the simulation of this model on
 a Barab\'asi-Albert (BA) network and a Watts-Strogatz(WS)
 network\cite{Garlaschelli2004, Souma2001}.
 In these studies, the authors  discussed   numerical
 simulation results, however, none of them  proposed a quantitative theory
 that could  explain these  results.
 
 This study aims  to develop a quantitative theory of the BM model on a
 random  network.
 The key assumption of our theory is   "adiabatic and independent"
 assumptions on the stationary distribution function, which is
 explained  in a  later section. By using these assumptions and the central
 limit theorem,
 we analytically derive the equations that determine the  stationary
 distribution function in the non-wealth-condensate phase. We
 compared our analytic  results  with those of the numerical
 simulation, and we found that our 
 theory showed better agreement than did the  mean-field theory.

 The remainder of this paper is organized as follows. In the next
 section, we define the
 model we investigate in this paper. Then, we describe our theory and
 its results, first for the case of a  regular random network and then
 for a random network with arbitrary degree distribution. In Sec. 
\ref{084157_4Apr12},  we describe a comparison of 
 our analysis and the numerical simulation. In the last section, we
 summarize our results and discuss the problem to be solved.

\section{Model}

The original BM model is expressed using  the  Stratonovich stochastic
differential equation; however, we use the equivalent Ito stochastic
equation for mathematical convenience. 
We consider the BM model on a complex network described by the following Ito
stochastic differential equations
\begin{equation}
{dx_i}= [J \sum_{j=1}^N a_{ij}(x_j-x_i)]dt + \sqrt 2 \sigma x_i dW_i,\label{BM-eq}
\end{equation}
where $x_i$, $J$, $N$, and $\sigma^2$  are the same as
 Eq.(\ref{BM-original}),
 and $a_{ij}$ represents the adjacent matrix.  On the
network model, we consider a random network in which the degree distribution
is given by $Q(k)$.

\section{Theory}
\label{173945_16Apr12}

 In this  section, we consider the stationary PDF of the normalized wealth of
 Eq.(\ref{BM-eq}). 
 Eq.(\ref{BM-eq}) is invariant
 under the change of
 scale  $x_i^{\prime}=\alpha x_i$ for any positive constant $\alpha$, and we can assume
 $\langle x \rangle = 1$ without loss of generality.
We derive the analytic form of the stationary PDF of
 $x_i/\langle x \rangle$, the normalized
wealth at node $i$. First, we explain our method on the
regular random graph, in which all nodes have the same degree $k$. Then,
 we extend the analysis to the general random network model, whose
 degree distribution is given by $Q(k)$.

\subsection{Case of a regular random network}

 This subsection focuses on the analysis of the system when  each
 node has the same degree $k$, in other words, $Q(k)$ is a delta function.
 We assume that $\rho_i(x,t)$, the PDF  of wealth at node $i$,  is
 independent of  $i$,
 $\rho_i(x,t) = \rho(x,t)$.

First,  we review the mean-field treatment of the
BM model. We consider  the distribution of wealth at  node $i$. By using
the mean-field theory, we approximate
$\frac{1}{k}\sum_j a_{ij}x_j =\langle x \rangle= 1$ in
Eq. (\ref{BM-eq}). Under this approximation, $\rho(x,t)$ satisfies the
following Fokker-Planck equation,
\begin{equation}
\frac{\partial \rho }{\partial t} = -\frac{\partial}{\partial x}\left[(J
								 k (1-x)
								 -\sigma^2
								 x) \rho\right] +\sigma^2 \frac{\partial}{\partial x}
\left[x \frac{\partial}{\partial x} (\rho x)\right] \label{BM-MF-eq}
\end{equation}
and we find $\rho^{MF}_{eq}(x)$, the stationary PDF obtained from this
  equation,  as
\begin{equation}
\rho^{MF}_{eq}(x)= C \exp(-\alpha/x) x^{-2-\alpha},
\end{equation}
where $\alpha= J k /\sigma^2$ and $C=\alpha^{1+\alpha}/\Gamma(1+\alpha)$.
In this case, wealth-condensation, defined as the divergence of
$\langle x^2\rangle$, occurs at $J k  \le \sigma^2$. 

 To proceed  beyond the mean-field approximation, we make 
 what   we call ``adiabatic and independent'' assumptions. We
define the "local"
 field $\tilde x_i = \frac{1}{k}\sum_j a_{ij} x_j$.
 Because $a_{ij}$ is not
 0 only if nodes $i$ and $j$ are connected, $\tilde x_i$ represents the
 local average of the wealth  around node $i$. 

 If $\tilde x_i$ is fixed, the PDF of $x$ at node $i$ is given by solving
 the "local" Fokker-Planck equation
\begin{equation}
\frac{\partial \rho }{\partial t} = -\frac{\partial}{\partial x}\left[J k (\tilde x_i -x)
 \rho  -\sigma^2 x \rho \right] +\sigma^2 \frac{\partial}{\partial x}
\left[x \frac{\partial}{\partial x} (\rho x)\right],\label{localFP}
\end{equation} 
and the  conditional PDF of $x$ under the local field $\tilde x$ is given by
\begin{equation}
\rho_{eq}(x|\tilde x) = C(\tilde x) \exp(-\alpha \tilde x/x)x^{-2-\alpha},\label{condPDF}
\end{equation}
where $C(\tilde x)= (\alpha \tilde x)^{1+\alpha}/\Gamma(1+\alpha)$.

Here we make  the ``adiabatic and independent'' assumptions.
 First, we assume the static PDF  $\rho_{eq}(x)$ can be approximated
 by
\begin{equation}
\rho_{eq}(x) = \int d\tilde x P(\tilde x)\rho_i(x| \tilde x),\label{staticAssume}
\end{equation}
where $P(\tilde x)$ is the PDF of
$\tilde x$.  If $\tilde x$ changes much slower than  $x$,
 this condition is satisfied, and therefore  we call it the
 ``adiabatic'' assumption. 
 Under this assumption,
the problem to calculate $\rho_{eq}(x)$ is reduced to the one to calculate
$P(\tilde x)$.

The second assumption is needed to calculate $P(\tilde x)$.
We assume that the random variables $x_j$, where $j$ runs in
the neighborhood  of node $i$,  are independent. This assumption enables
us to use the central limit theorem to obtain  $P(\tilde x)$.
 We consider the case in which  the variance of $x$ is   finite, and that
 the average and variance of $x$ are 1 and $ s^2$,  respectively.
 Using the central limit theorem, we can approximate
\begin{equation}
P(\tilde x)\sim\frac{\sqrt k}{\sqrt { 2\pi} s}\exp\left(-\frac{k (\tilde x - 1)^2}{2 s^2}\right).\label{142356_4Apr12}
\end{equation}
Inserting  Eq. (\ref{142356_4Apr12}) into Eq. (\ref{staticAssume}), we obtain
\begin{equation}
\rho_{eq}(x) \sim \int_{0}^{\infty} d\tilde x \frac{\sqrt k}{\sqrt{2 \pi}s} \exp\left(-\frac{k (\tilde x -1)^2}{2 s^2} \right) \rho_{eq}(x|\tilde x).\label{total-dist-eq}
\end{equation}

The final step is to check that $\langle x \rangle=1$ and to calculate $s^2$.
 Using $\int_0^{\infty} dx \rho_{eq}(x|\tilde x) x =
\tilde x$ and $\int_0^{\infty} dx \rho_{eq}(x|\tilde x) x^2  =
\frac{\alpha}{\alpha-1}\tilde x^2$ for $\alpha > 1 $,  we obtain
\begin{equation}
 \int x\rho_{eq} (x) = \int_0^{\infty} d\tilde x P(\tilde x)
  \tilde x \sim \int_{-\infty}^{\infty} \tilde x P(\tilde
  x)d\tilde x = 1,\label{101611_13Apr12}
\end{equation}
 and 
\begin{equation}
 \int x^2 \rho_{eq} = \int_0^{\infty} d\tilde x
  \frac{\alpha}{\alpha-1} \tilde x^2 P(\tilde x) \sim
  \frac{\alpha}{\alpha-1} \left(1+\frac{s^2}{k}\right).\label{101846_13Apr12}
\end{equation}
Here we change the lower limit of integration from 0 to $-\infty$,
assuming $k/s^2$ to be large.
From Eq.(\ref{101611_13Apr12}),  we conclude  that
 $\langle x \rangle =1$ , and that  there is no inconsistency. From Eqs. 
(\ref{101611_13Apr12})
and (\ref{101846_13Apr12}), the variation of $x$ is given by
$\frac{1}{\alpha-1}(\frac{\alpha s^2}{k}+1)$. Therefore we find the
self-consistency condition for $s^2$ as 
\begin{equation}
[(k-1)\alpha -k]s^2 = k. \label{eq_for_s}
\end{equation}

Eqs.(\ref{condPDF}), (\ref{total-dist-eq}), and (\ref{eq_for_s}) are the
set of equations that determine $\rho_{eq}(x)$.

We should make several comments on these results.
First, this theory gives a distribution that does not obey the
power-law behavior. Eq.(\ref{total-dist-eq}) shows that the PDF
 $\rho_{eq}(x)$
is written with the integral  of $P(\tilde x)\rho_{eq}(x|\tilde x)$ over
$\tilde x$. Because $P(\tilde x)$ is Gaussian, $\rho_{eq}(x)$ does not
exhibit the simple power-law behavior. 

The next important suggestion of  this theory concerns the
wealth-condensation transition. In our theory, $s^2$ diverges if
$(k-1)\alpha -k = 0$,  implying that the wealth-condensation occurs at this
point.  Using $\alpha=Jk/\sigma^2$, this leads to the conclusion that wealth
 condensation occurs at $J_c =\sigma^2/(k-1)$. On the other hand,
 the mean-field theory gives the divergence of the variance at
 $J_c= \sigma^2/k $.
 Therefore the difference between our theory and the mean-field theory can be
 tested by estimating  $J_c$ from the simulation. 

 Finally, we note that our theory can be
 applied only for the non-wealth-condensation phase. We need the central
 limit theorem to obtain $P(\tilde x)$, which is only applicable for the
 case in which the variance of $x$ is finite. We discuss  this point in
 greater detail in the final section.

\subsection {Case of a general random network}

 In this subsection, we extend the developed method for a regular
 random graph to a general random network.

 As in the case of a regular random network, we start from the mean-field
 theory. In this theory, we define
 $\rho_k(x,t)$ as the PDF of $x$ on the node whose degree is $k$.
 Because this network is heterogeneous, the mean
 of $x$ may depend on $k$. Therefore, we need to define
 $\bar x_k = \int dx x \rho_k(x,t) $, the average of $x$ on nodes with
 degree $k$,  to perform the mean-field calculation.
 The mean-field Fokker-Planck equation is constructed  in the same
 manner as in the case of the  SIS model or Kuramoto
 oscillator\cite{Pastor-Satorras2001, Ichinomiya2004}, and we obtain
\begin{eqnarray}
 \frac{\partial \rho_k(x,t)}{\partial t } & = &  -\frac{\partial}{\partial
  x}\left[J\sum_{k^{\prime}} \frac{kk^{\prime} Q(k^{\prime})}{\langle k
     \rangle}(\bar x_{k^{\prime}}-x)\rho_k(x,t) \right.\nonumber\\
 &-& \left.\sigma^2 x \rho_k(x,t)\right] +
 \sigma^2\frac{\partial}{\partial x}\left[x\frac{\partial}{\partial x }(x\rho_k(x,t))\right]\nonumber\\
\end{eqnarray}
 We can simplify this equation by introducing the ``weighted'' average of
  $x$ as  $\bar x = \sum_k \frac{kQ(k)}{\langle k \rangle} \bar x_k$, which leads to 
\begin{equation}
 \frac{\partial\rho_k}{\partial t}= -\frac{\partial}{\partial x
  }[Jk(\bar x -x)\rho_k -\sigma^2 x \rho_k] +
  \sigma^2 \frac{\partial}{\partial
  x}\left[x\frac{\partial}{\partial x }(x\rho_k)\right],
\end{equation}
where we abbreviated $\rho_k(x,t)$ as $\rho_k$.
As  in the  case of a regular random graph, we obtain
 $\rho^{MF}_{eq,k}(x)$, the stationary PDF by the mean-field theory, as 
\begin{equation}
 \rho^{MF}_{eq,k}(x)=C_k(\bar x) \exp(-\alpha_k\bar x/x)x^{-2-\alpha_k},\label{091511_4Apr12}
\end{equation}
where  $\alpha_k = Jk/\sigma^2$ and $C_k(\bar x)=(\alpha_k \bar
x)^{1+\alpha_k}/\Gamma(1+\alpha_k)$. Finally, $\bar x$ is determined to satisfy
the condition $\langle x \rangle = 1$.
From $\int dx x \rho^{MF}_{eq,k}(x) = \bar x$, we obtain  $\bar x = 1$.

 Now we follow the same procedure as that described  in the previous
 subsection to calculate the PDF more accurately. We consider the PDF of
 the  wealth on node $i$,  whose 
 degree is  $k$. If the average of sums of $x$ in the neighborhood of node
 $i$ is given by $\tilde x$, we find the conditional PDF of $x$  on node $i$
 as
\begin{equation}
 \rho_k(x|\tilde x) = C_k(\tilde x)
  \exp(-\alpha_k\tilde x/x)x^{-2-\alpha_k},\label{generalrhop}
\end{equation}

Using the adiabatic assumption explained in the previous subsection, we 
assume
\begin{equation}
 \rho_k(x) = \int d\tilde x P_k(\tilde x) \rho_k(x|\tilde x).\label{generalrho}
\end{equation}

Next, we approximate $P_k(\tilde x)$ by Gaussian using the independent
 assumption. One difference between the regular random graph and the random
 graph with arbitrary degree distribution lies in that we need to assume
 Lindeberg's  condition in this case.
Suppose that  there are independent variables  $y_{1},\cdots y_{m}$ , whose
 mean and variance  are $\mu_1, \mu_2, \cdots \mu_m$ and
  $s_1^2, s_2^2, \cdots s_m^2$ respectively. Then, the PDF of 
 $\frac{1}{m}(y_1+y_2 +\cdots y_m)$ converges to the  Gaussian, whose mean and
 variance are $\frac{1}{m}(\mu_1+\mu_2 +\cdots +\mu_m)$ and $\frac{1}{m^2}
(\sigma_1^2 + \sigma_2^2 + \cdots \sigma_m^2)$,  for $m\rightarrow
 \infty$, if Lindeberg's condition is satisfied\cite{Gnedenko}.
 Although $m$ is finite in our case, we can approximate
\begin{equation}
 P_k(\tilde x)\sim \frac{1}{\sqrt{2\pi} S_k} \exp\left(-\frac{(\tilde x
					   -1)^2}{2
					   S_k^2}\right)\label{generalp},
\end{equation}
where $S_k^2$ is a parameter that must be determined by the self-consistency
 condition. 

The final step is to obtain self-consistent equations for the mean and
variance of $x_k$.
Using
$\int dx x \rho_k(x) = 1$ and $\int dx x^2 \rho_k(x) =
\frac{\alpha_k}{\alpha_k-1}(1+ S_k^2)$ obtained from
 Eqs.(\ref{generalrhop}), (\ref{generalrho}) and (\ref{generalp}),
 we respectively obtain  $\mu_k$ and $s_k^2$, the average and 
 variance  of $x$ on
 nodes with degree $k$, as $\mu_k=1$ and 
\begin{equation}
s_k^2 =\frac{\alpha_k S_k^2 +1}{\alpha_k-1}.\label{111851_13Apr12}
\end{equation}

  A node with degree $k$ is connected to a node whose  degree is
  $k^{\prime}$ with probability $\frac{k k^{\prime } Q(k^{\prime})}{\langle k
\rangle}$, and  we obtain
\begin{equation}
 S_k = \frac{1}{k^2}\sum_{k^{\prime}} \frac{kk^{\prime}Q(k^\prime)}{\langle k
  \rangle} {s_{k^{\prime}}}^2 = \frac{u}{k},\label{generalS}
\end{equation}
where $u$ represents the weighted average of the variance,
\begin{equation}
 u=\sum_k kQ(k)s_k^2/\langle k \rangle.\label{112117_13Apr12}
\end{equation} 
From Eqs.(\ref{111851_13Apr12}) and (\ref{generalS}), we obtain
\begin{equation}
 s_k^2 = \frac{\alpha_k u/k+1}{\alpha_k -1}.\label{generals}
\end{equation}
Inserting this equation into Eq.(\ref{112117_13Apr12}), we obtain
\begin{equation}
 u = \sum_k \frac{Q(k)}{\langle k \rangle}\frac{\alpha_k u +k}{\alpha_k-1},
\end{equation}
which leads to our final result,
\begin{equation}
 \left(1-\sum_k\frac{Q(k)\alpha_k}{\langle k
  \rangle(\alpha_k-1)}\right) u =\sum_k\frac{Q(k)k}{\langle k \rangle(
 \alpha_k-1)}\label{generalu}
\end{equation}
Eqs.(\ref{generalrhop}), (\ref{generalrho}), (\ref{generalp}),
(\ref{generalS}),  and (\ref{generalu}) are the set of 
equations that  determine the stationary PDF.

The most important difference between the mean-field theory  and
ours lies in  that there is no effect of  $Q(k)$ in the
mean-field theory.
In the mean-field theory,  the PDF on the node with degree $k$ given
by Eq.(\ref{091511_4Apr12}) does
not depend on $Q(k)$, which  causes a  curious behavior. Suppose that we 
 decrease $J$ from a large value to 0. At
$J=\sigma^2/k$, the node with degree $k$ goes into the wealth-condensate
phase, whereas nodes with a larger degree do not
condensate. Therefore, the  mean-field theory predicts the coexistence of 
condensated and non-condensated nodes. In this theory, low-degree
nodes condensate at large $J$, whereas high-degree nodes do not
condensate until $J$ becomes sufficiently small.

On the other hand, all PDFs with different degrees are connected through
$u$ in our theory. In this theory, $u$ diverges when $\sum_k
\frac{Q(k)\alpha_k}{\langle k \rangle (\alpha_k-1)}= 1$, which implies
the divergence of  all $s_k$ from Eq.(\ref{generals}). In our theory,
  wealth-condensation occurs on all nodes simultaneously.

 Finally, we  comment on the behavior
 of the BM model  on a scale-free network.
 Researchers in the field of  complex networks may consider that
 a  singular behavior occurs in a scale-free network upon the 
 wealth-condensation transition, such as
 the divergence or disappearance of the transition point $J_c$.
 Unfortunately,this is not the case.
 The main difference between our case and other models that show
 singularity, such as the SIS-model or Kuramoto transition, lies in that
 we impose the  self-consistency condition on the variance of $x$,
 and not on its average. As shown in Eq.(\ref{generalS}), we divide the
 weighted average
 of the variance $s_k^2$ by $k$ to calculate $S_k$. This eliminates the
 singular behavior that  we often observe in the dynamics on a scale-free
 network.

\section{Simulation}
\label{084157_4Apr12}
 In this section, we test the analytic results obtained in the
 previous section by comparing them with numerical simulations.

 Because we have presented  two analytic results, one  for a regular
 random network
 and the other for a heterogeneous random network, 
 we carried out the simulations for both networks.
 For the former, all nodes had the  same degree, whereas for the latter,
  half of the nodes had degree 10 and the other half had degree 20.

 We use these two models because it is easy to calculate the PDF in
 these models. 
 However, the readers might find  these models too
 artificial.  To test our theory in a more realistic model, we
 show the result of simulations on BA-network\cite{Barabasi1999}.
 Here we note that the direct calculation of wealth distribution is
 rather  hard because we must calculate integral in
 Eq.(\ref{generalrho}) for many values of $k$.
 In this paper,  we investigate $s_k^2$, the variation of $x$ at a node
 with degree $k$ to test our theory instead, because it is much easier to
 compute.

 In the following simulation,  we carried out numerical integration
 by the Euler-Maruyama algorithm to obtain
 the distribution of normalized wealth $x/\langle x \rangle$.

\subsection{Case of a regular random network}

 In this subsection,  we show the  simulation result for a  regular random
 network and compare it with our theory.

\begin{figure}[t]
\resizebox{.3\textwidth}{!}{\includegraphics{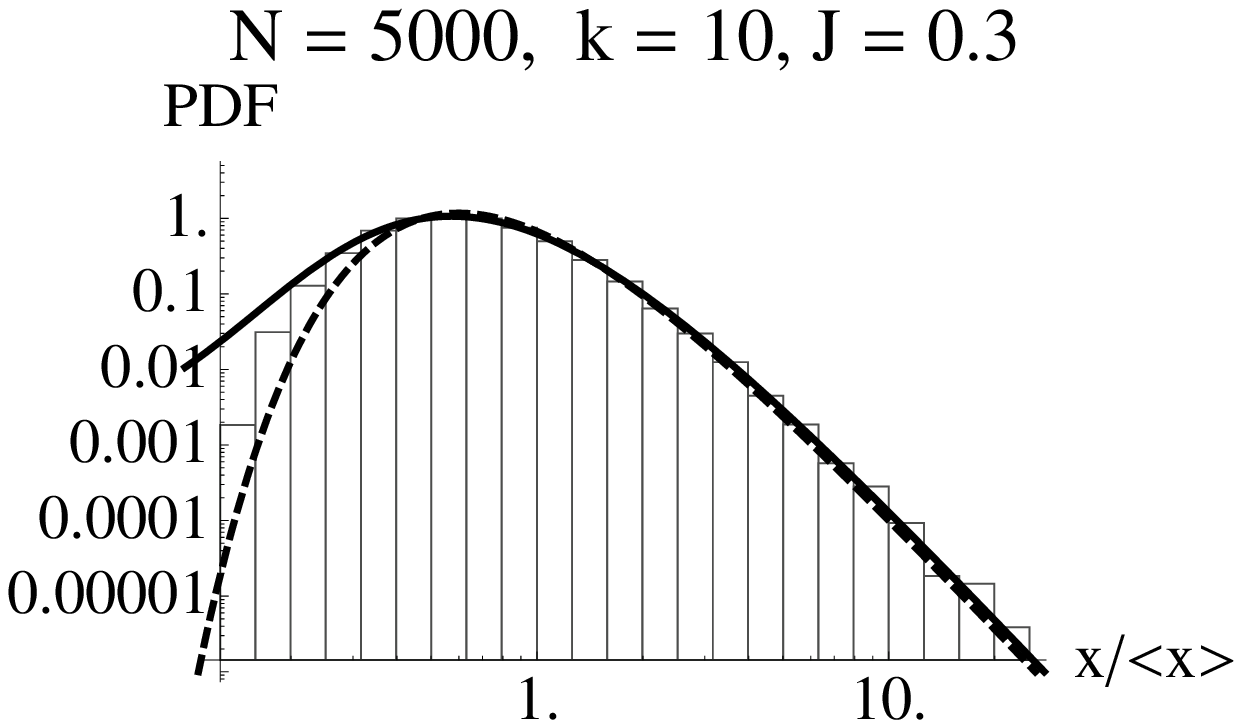}}
\resizebox{.3\textwidth}{!}{\includegraphics{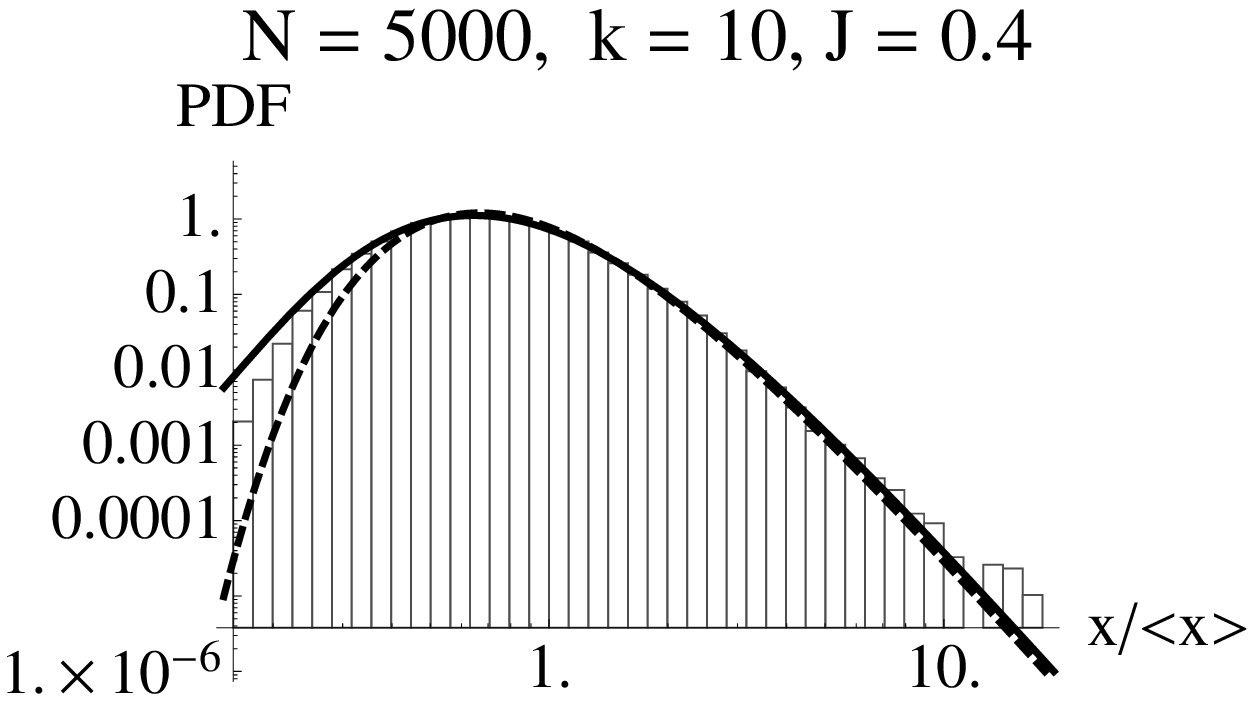}}
\resizebox{.3\textwidth}{!}{\includegraphics{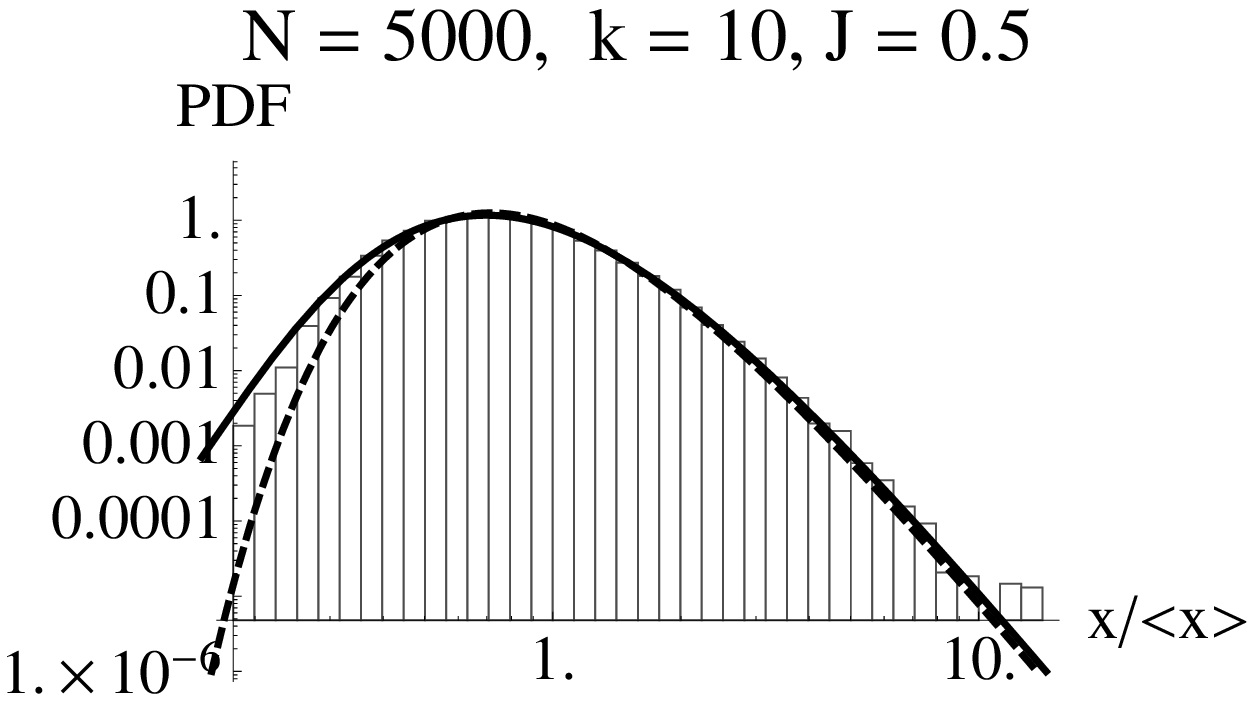}}
\caption{ Log-log plot of PDF of normalized wealth obtained by
 simulation when $k=10$, $\sigma^2=1$, and $J=0.3, 0.4$, and 0.5. The
 solid and dashed lines indicate the result obtained by our theory and
 that obtained by the mean-field theory,respectively. \label{d10fig}} 
\end{figure}

 First we calculate the PDF of the normalized wealth from 10 simulation
 trials on networks that include 5000 nodes. 
 In Fig.\ref{d10fig},
 we show the  obtained PDF for  the network with degree $k=10$,
 $\sigma^2=1$ and coupling $J=0.3$, 0.4, and 0.5.
 For all $J$, the distribution  coincides well  with our theory,
 indicated by the solid line. 
 The PDF given by the mean-field theory, indicated by the dashed line,
 strongly underestimates the probability density at small
 $x/\langle x \rangle$. On the other hand, our theory gives a slightly
 larger PDF than the simulation, although, it shows better agreement.

\begin{figure}[t]
 \resizebox{.45\textwidth}{!}{\includegraphics{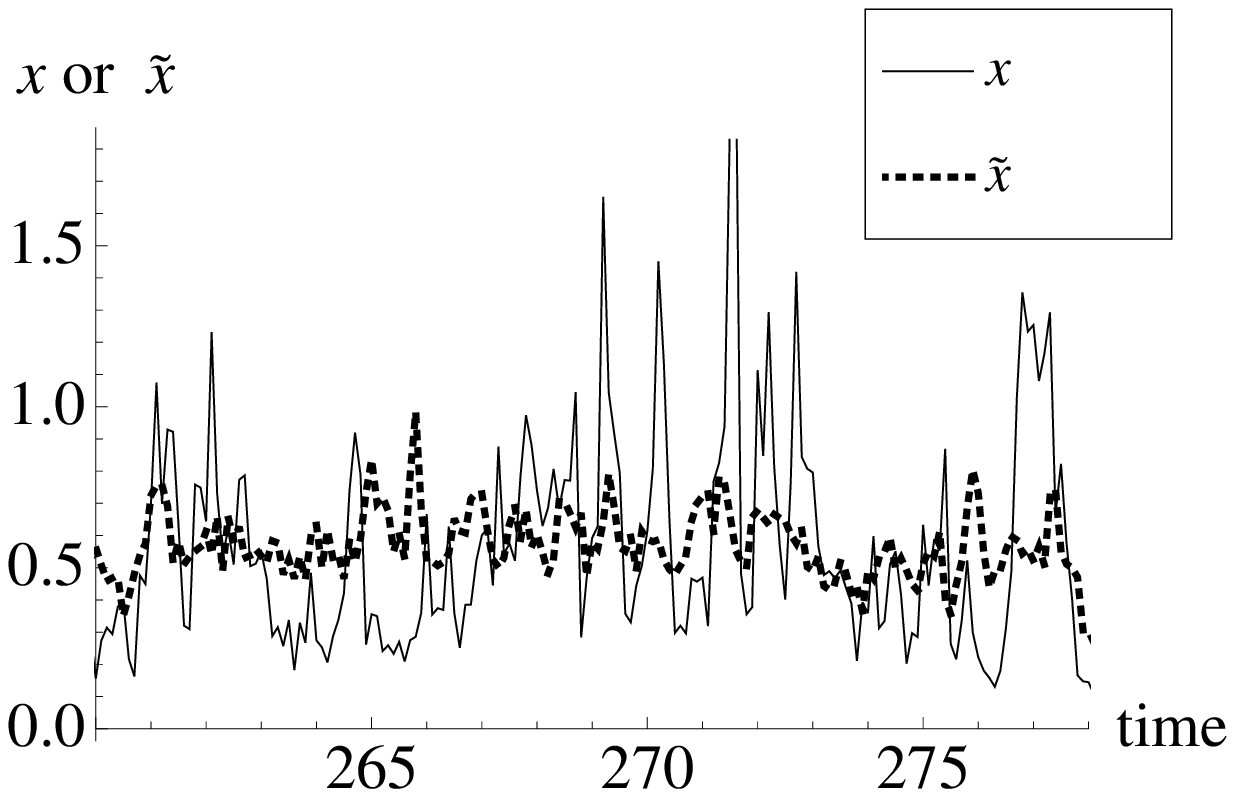}}
 \resizebox{.45\textwidth}{!}{\includegraphics{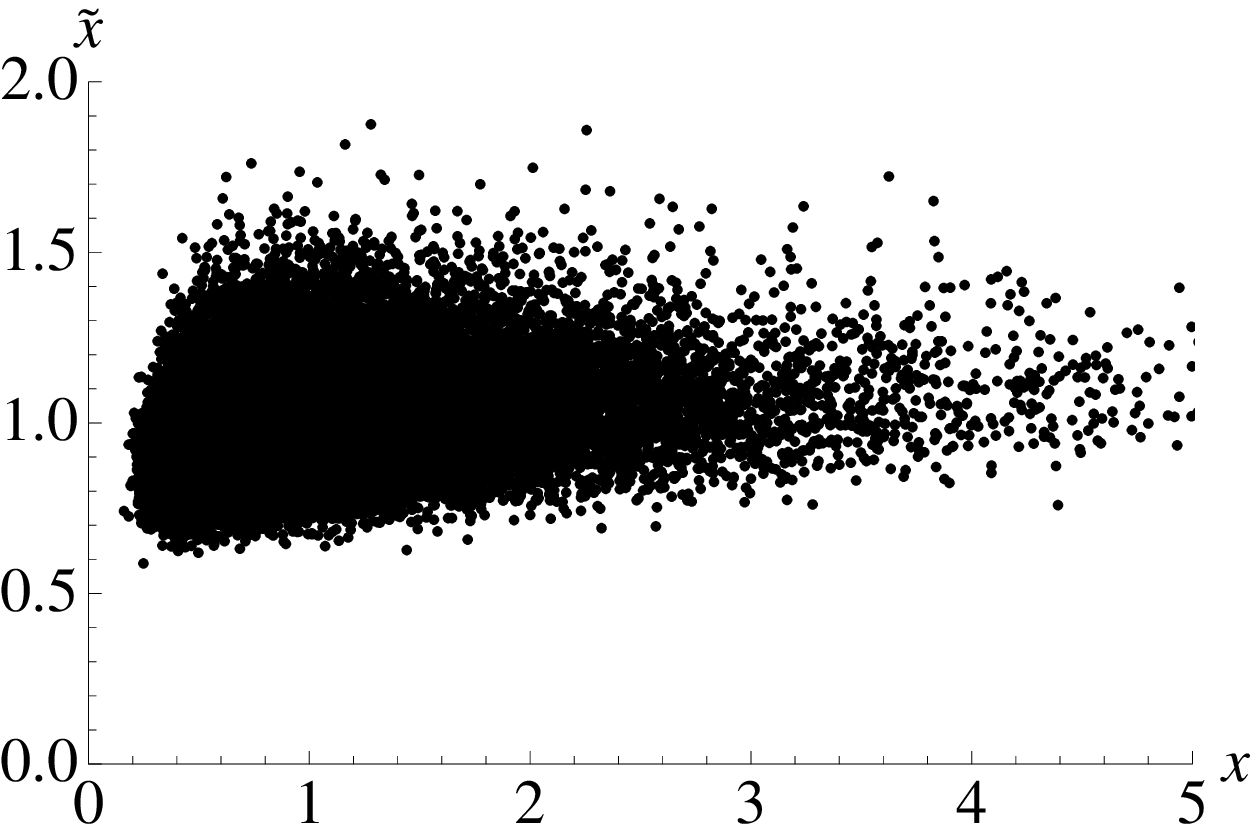}}
\caption{Temporal and spatial behavior of  $x$. Left: time
 series of $x$ and $\tilde x$ at one node. Right: scatter plot of $x$
 and $\tilde{x}$. Both plots are  obtained  from the simulation at $N=5000$,
 $k=10$ and $J=0.3$.}
\label{173853_3Aug12}
\end{figure}

 Here we investigate the temporal and spatial correlation of  $x$
 to test the ``adiabatic and independent''
 assumptions used to derive Eqs.(\ref{staticAssume}) and
 (\ref{total-dist-eq}). In the right-hand
 side of  Fig.\ref{173853_3Aug12}, we plot the typical trajectory of 
$x(t)$ and $\tilde x(t)$. Because
 the model we study is described by stochastic differential equations,
 it is difficult to show  that the change in $\tilde{x}_i$ is ``slower''
 than that in $x_i$,
 however, it seems that  $x_i$ changes more quickly than $\tilde x_i$. 
 Concerning the ``independent'' assumption, 
 we show the scatter plot between $x$ and $\tilde x$ in the left-hand
 side of Fig.\ref{173853_3Aug12}. Although there exists a tendency for 
 $x$ to increases as  $\tilde x$ increases, it is not strong.
 The correlation between $x$ and $\tilde x$ calculated by the 
 numerical simulation is 0.238. Therefore we conclude that the ``adiabatic''
 ``independent'' assumptions are fairly good.

\begin{figure}[t]
\resizebox{.3\textwidth}{!}{\includegraphics{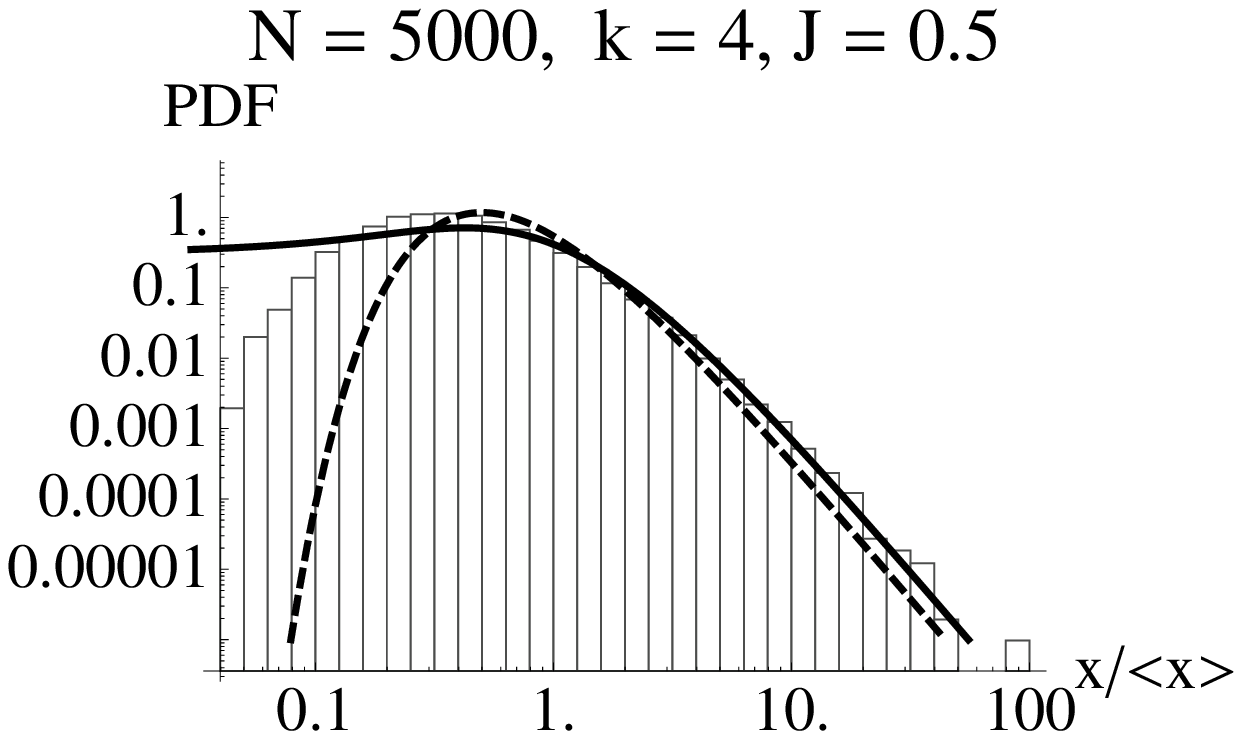}}
\resizebox{.3\textwidth}{!}{\includegraphics{N5000_k10_J05.eps}}
\resizebox{.3\textwidth}{!}{\includegraphics{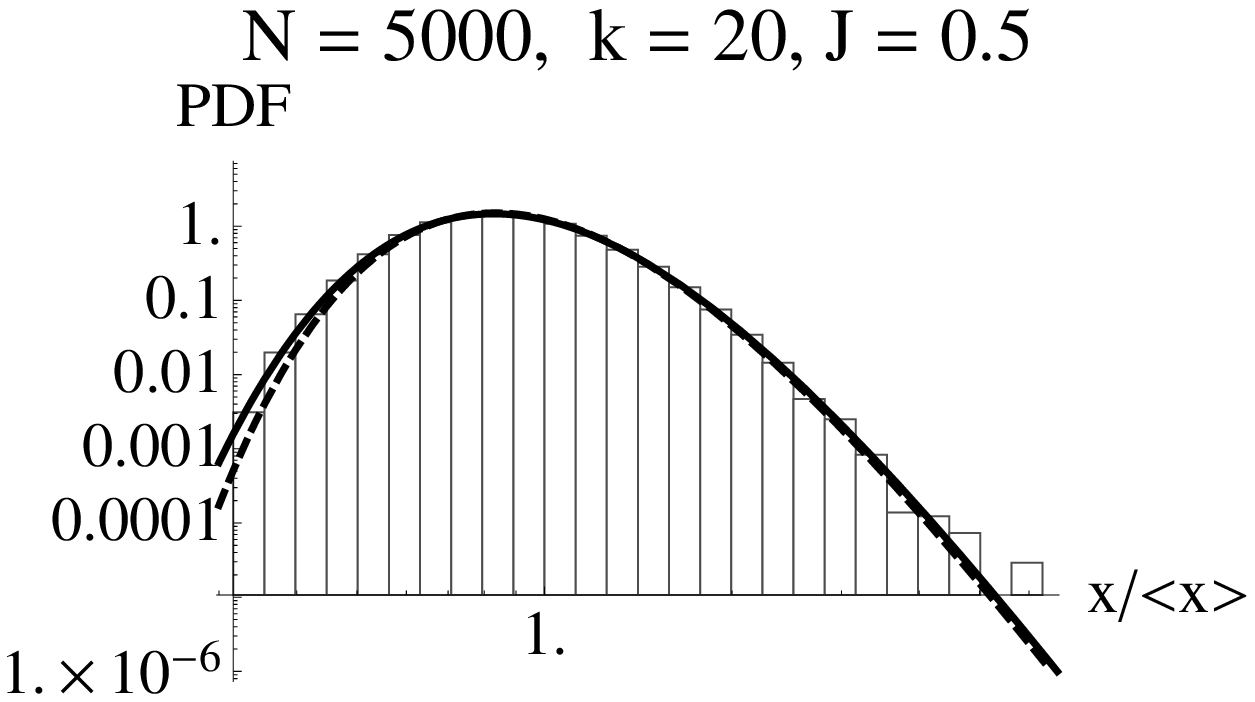}}
\caption{ Log-log plot of PDF obtained by
 simulation when $J=0.5$, $\sigma^2=1$ and $k=4, 10$, and 20. The solid
 and dashed lines indicate  the result obtained by our theory and that
obtained  by the  mean-field theory, respectively \label{J05fig}.} 
\end{figure}

 In Fig. \ref{J05fig}, we show the PDF at $J=0.5$, $\sigma^2 = 1$ and
 $k=4, 10$,  and 20. It is clear that the
 discrepancy between our theory and the numerical simulation increases
 as $k$ 
decreases. This is not surprising, because the error caused by
 the application of the central limit theorem increases at small
 $k$.
 However, our theory appears to be  better than the   
 mean-field theory, even in the worst case. For example, when $k=4$,
 the probability density obtained by the mean-field
 theory is smaller than $10^{-5}$ at $x=0.1$, whereas the  numerical simulation
 and our theory shows that it is $O(10^{-1})$. On the other hand, the
 difference from the result of numerical simulation  at large $x$ is
 indistinguishable between our theory and mean-field theory. 

\begin{figure}[t]
\resizebox{.4\textwidth}{!}{\includegraphics{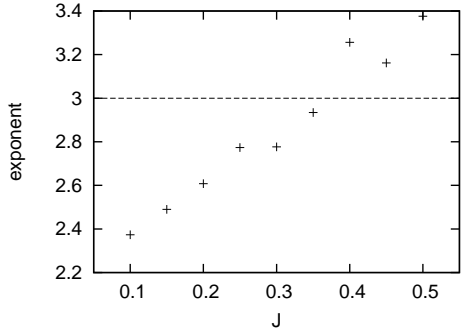}}
\caption{ Exponent obtained by simulation for  $k=4$, $\sigma^2 =
 1$. The dotted line indicates the wealth-condensation transition.\label{variance}}
\end{figure}

 Finally, we check the wealth-condensation transitions. Though our
 theory does not exhibit the exact power-law behavior, it suggests that
 variance of normalized wealth diverges at $J_c=\sigma^2/(k-1)$.
 Therefore, we can expect power law behavior $\rho(x)\propto x^{-3}$ for
 large $x$ at this point. On the other hand, the mean-field
 approximation gives $J_c=\sigma^2/k$, and we can test our theory by
 estimating $J_c$ by checking the exponential tail.  In
 Fig.\ref{variance} we plot the exponent $\gamma$, $\rho(x)\propto
 x^{-\gamma}$ at $x\gg 1$, estimated  from the
 fitting of the PDF obtained  through 100 simulation trials  on a regular random
 graph $N=1000$, $k=4$. 
 It is clear that 
 wealth-condensation occurs at $J\sim 0.35$, which is closer to the
 prediction of our theory $J_c=0.33$, than that of the mean-field theory
, $J_c=0.25$.

\subsection{Case of a heterogeneous network}
 In this section, we present the  simulation result for a  heterogeneous 
 network. As explained in Sec.\ref{173945_16Apr12}, the PDF
 $\rho_{eq,k}(x)$ does not depend on  $Q(k)$ in the mean-field
 approximation. In this approximation, the PDF of the wealth for nodes whose
 degree is $k$ equals  that for the regular random graph. On the other
 hand, our theory predicts that  the PDF changes if $Q(k)$
 differs. 
 To test our theory for heterogeneous network, we make the
 network that half of the node have degree $k_1=10$, and other half have
 degree $k_2=20$.

\begin{figure}[t]
 \resizebox{.45\textwidth}{!}{\includegraphics{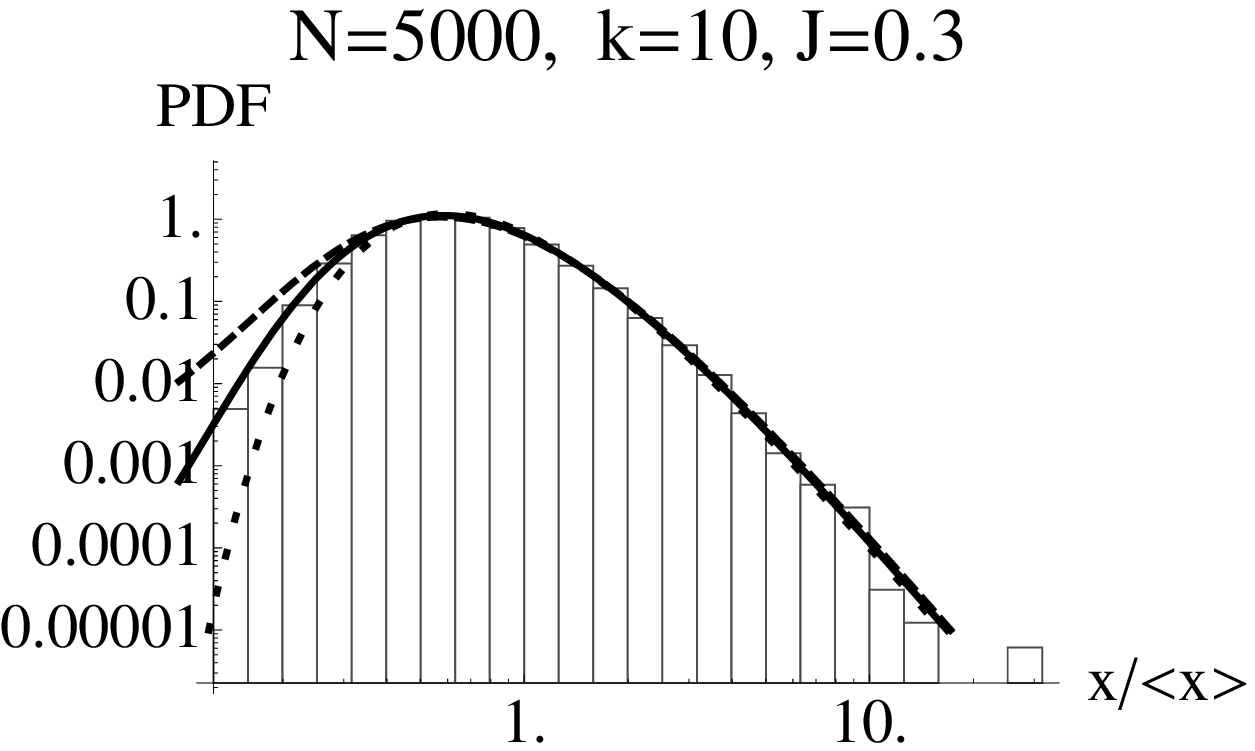}}
 \resizebox{.45\textwidth}{!}{\includegraphics{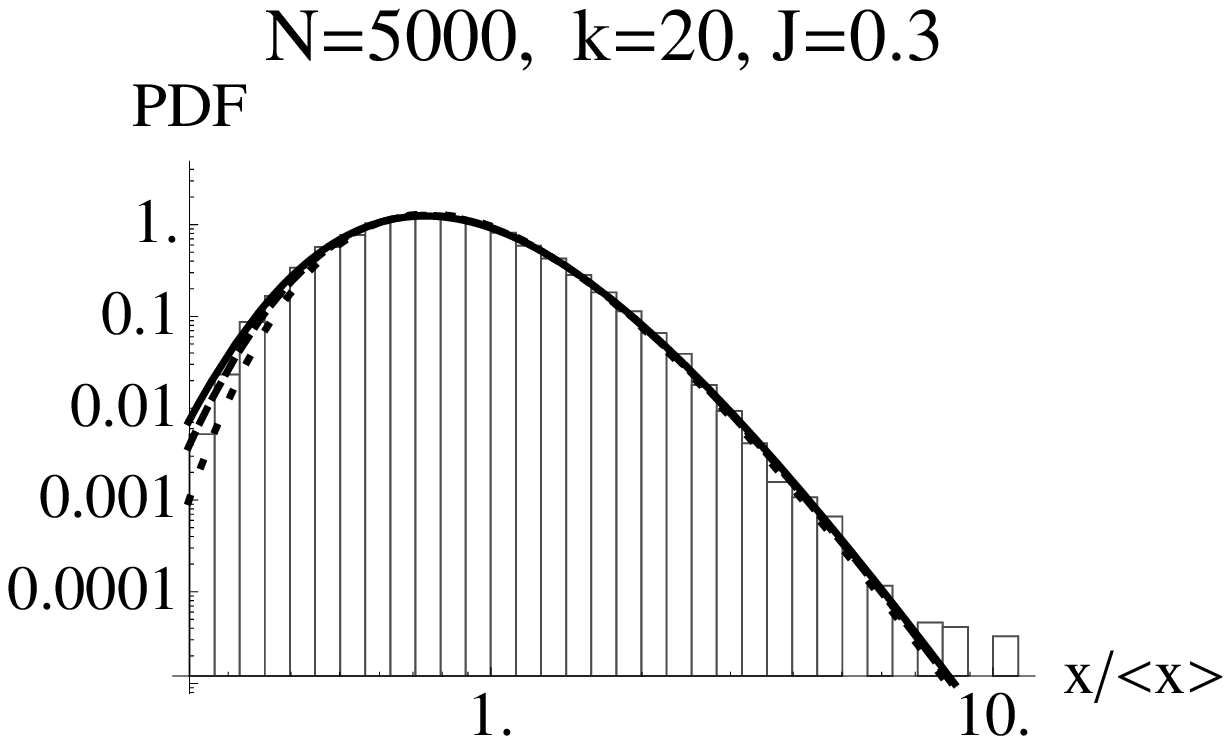}}
 \caption{ PDF obtained from numerical simulation on a heterogeneous
 network. The solid, dashed, and dotted lines indicate the PDF obtained
 by our theory for a heterogeneous network and a  regular random graph,
 and by the mean-field theory, respectively}
 \label{091054_12Apr12}
\end{figure}

 The PDF of the normalized wealth obtained by the simulation is shown in
 Fig.\ref{091054_12Apr12} for $J=0.3$ and $\sigma^2 = 1$. In this
 figure, we also plot the PDF obtained by our theory for a heterogeneous
 network and a regular random graph with degree  $k_1$ or $k_2$ and 
 by the mean-field theory and that by our theory for regular random graph
 with degree $k_1$ or $k_2$. For the PDF on
 nodes with degree 10, the PDF obtained by our theory, indicated  by the
 solid line,  is suppressed  at small $x$ compared with the case of a 
 regular random network, indicated by the dashed line.
 The PDF obtained by the simulation shows better agreement with our
 theory for a heterogeneous network. For the PDF on node with degree 20, the
 difference among
 the three theories is small; however, we can conclude that our theory can
 estimate the PDF very well. 

\subsection{Case of BA-network}

\begin{figure}[t]
 \resizebox{.45\textwidth}{!}{\includegraphics{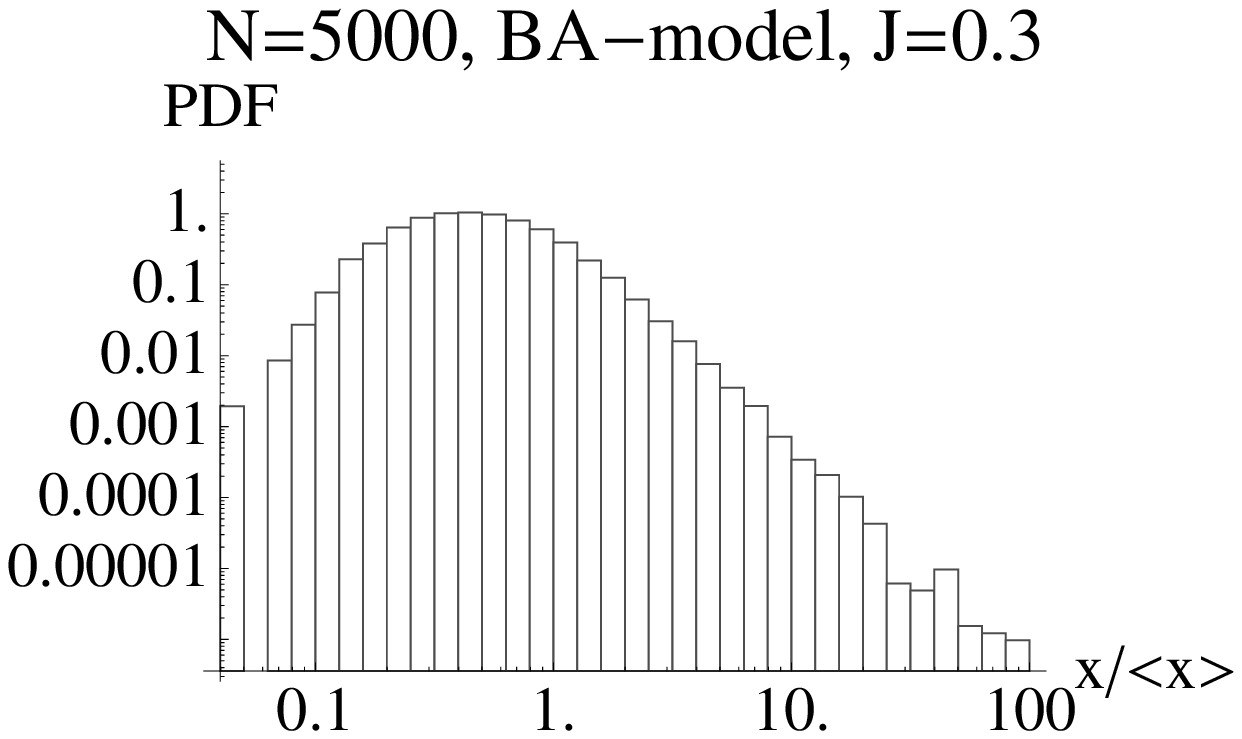}}
 \resizebox{.45\textwidth}{!}{\includegraphics{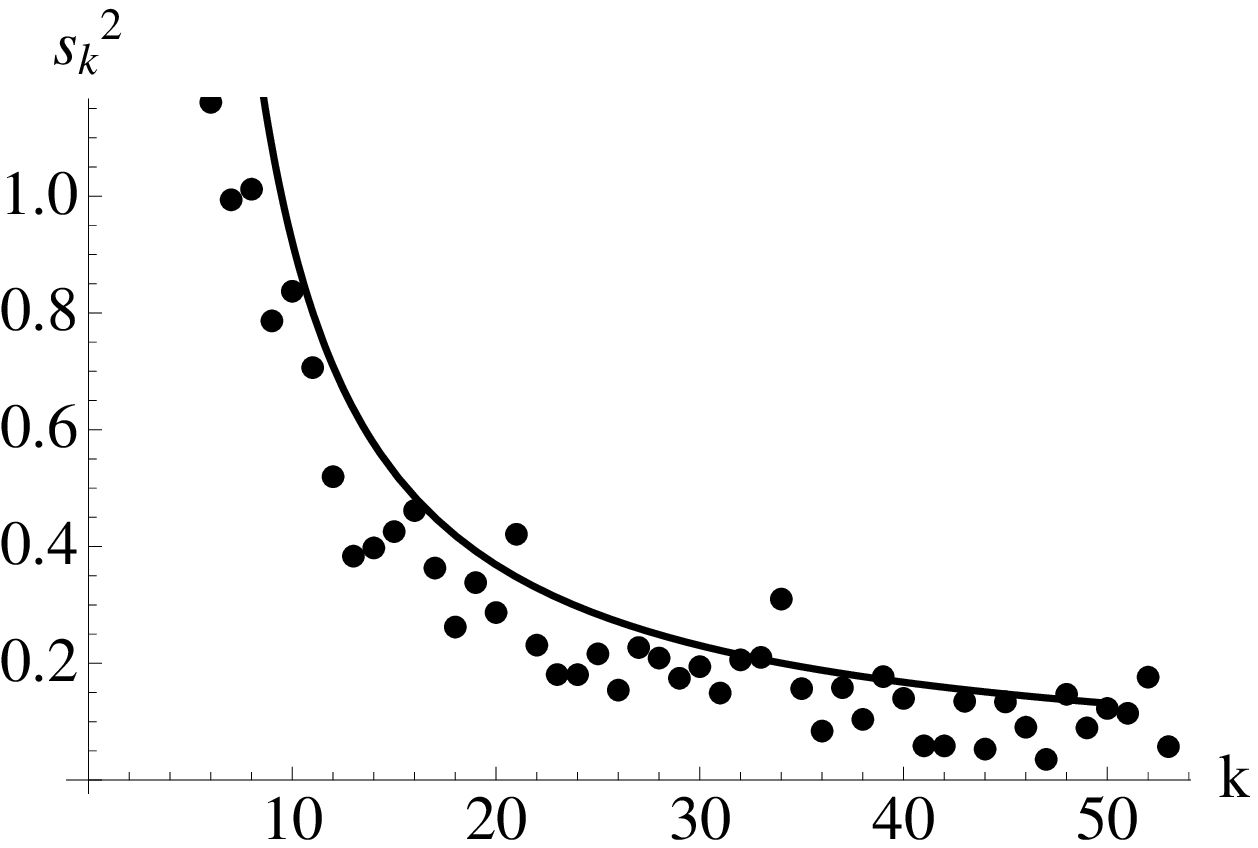}}
\caption{Result of simulation for BA-network with minimum degree 4
 and $J=0.3$, $\sigma^2=1.0$. Left: PDF. Right:
 $s_k^2$, the variation of $x$ on nodes with degree $k$. Thick line
indicates the theoretical result obtained from Eq.(\ref{generals}).}
\label{112226_3Aug12}
\end{figure}

 To test our theory in a more realistic network model, we  show the
 simulation result on the   BA-network model.
 The PDF obtained from  simulations on a
 BA-model with minimum degree 4,  $J=0.3$ and $\sigma^2 = 1$ is given
 in the left-hand side of  Fig.\ref{112226_3Aug12}.
 As we have noticed, it is slightly difficult to show the PDF obtained from
 our theory. Instead of the
 PDF, we plot the $s^2_k$, the variation of $x$ on nodes whose degree is $k$,
 obtained by simulation and that obtained by Eq. (\ref{generals}), in
 the  right-hand side of  Fig. \ref{112226_3Aug12} . The theoretical
 prediction, indicated by the  thick line, coincide well with the result
 of numerical simulation.

\section{Summary}

 In this paper, we derived the PDF of the 
 Bouchaud-M\'ezard model on a  random network. Using adiabatic and
 independent assumptions, we derived the equations for the PDF,
  Eqs. (\ref{condPDF}), (\ref{total-dist-eq}), and
 (\ref{eq_for_s}) for a regular random network, and
 Eqs.(\ref{generalrhop}), (\ref{generalrho}), (\ref{generalp}),
 (\ref{generalS}), and (\ref{generalu}) for a general random graph.
 It is difficult to solve these equations analytically; however, doing 
 so provides considerable  information about  the wealth-distribution.
 In particular,
 we can analytically obtain  the wealth-condensation point $J_c$  for a
 regular random graph.
 These analytic results are compared with the numerical simulations in
 Sec. \ref{084157_4Apr12}, and  good agreement is found with our theory.
 Below, we discuss the problem to be solved.

 First, we note that the approach we have used in this paper
 cannot be applied to a wealth-condensate phase. We use the
 central limit theorem to
 derive the equation for a stationary PDF; however,  this approach is not
 applicable when the variance of $\rho(x)$ diverges.
 To treat the wealth-condensate
 phase, we need a generalized central limit theorem that is  applicable
 when the variance diverges. It is known that the PDF of the sum of independent
 identical random variables converges  to a stable distribution
 function when the variance diverges\cite{Gnedenko}. It would be possible
 to make the theory for the 
 wealth-condensate phase on a regular random network using this theorem;
 however, further work will be required toward  this end. For a  general
 random network, the 
 generalized central limit theorem  unfortunately remains still
 insufficient. To treat such a network , we
 require a generalized central limit theorem for non-identical
 independent random variables, which has  not  been yet established.

 Second, one might be interested in   better approximating the
 the PDF. As shown in Fig. \ref{J05fig}, the PDF obtained by our method
 still has a  large discrepancy with the numerical simulation, especially for
 nodes with small degree.
 To reduce these errors, we need to use the improved central limit
 theorem, which can deal with the rate of convergence. The application of
 Chebyshev's rate-of-convergence theorem\cite{Chebyshev} would lead 
 to a  better approximation.

 Finally, we should  comment on the ``adiabatic and
 independent'' assumptions. We have no proof to verify these
 assumptions; however, the coincidence with the simulations suggests that these
 assumptions work very well. If so, the theory developed in
 this study  will be applicable to other dynamical systems. The essential
 part of our theory  is to  impose the self-consistency condition on the
 average and variance of the dynamical variables on each node. In the
 case of the BM
 model, the self-consistency condition for the average is automatically
 satisfied, and we need only to calculate the variance. 
 This procedure is very general, and it will be applicable to analyze other 
 dynamical behaviors such as synchronization or diffusion.

\begin{acknowledgments}
The author thanks H. Nakao, S. Morita, and T. Aoki for fruitful discussions.

This work is financially supported by PRESTO, Japan Science and
 Technology Agency.
\end{acknowledgments}


\begin{thebibliography}{99}
\bibitem{Pastor-Satorras2001}{ R. Pastor-Satorras and A. Vespignani,
	Phys. Rev. Lett. {\bf 86}, 3200(2001).}
\bibitem{Nishikawa2003} T. Nishikawa, A. E. Motter, Y.-C Lai, and
	F. C. Hoppensteadt, Phys. Rev. Lett. {\bf 91},014101(2003).
\bibitem{Ichinomiya2004} T. Ichinomiya, Phys. Rev. E {\bf 70} 026116(2004).
\bibitem{Nakao2008} H. Nakao and A. Mikhaikov, Nature Phys. {\bf 6} 544(2010).
\bibitem{Bouchaud2000} J. Bouchaud and M. M\'ezard, Physica A {\bf 282}
	536 (2000).
\bibitem{Pareto} V. Pareto, {\it Cours d'\'economie politique} Macmillan,
	London 1897.
\bibitem{Garlaschelli2004} D. Garlaschelli, M. I. Loffredo,
	Physica A. {\bf
	338} 113(2004).
\bibitem{Souma2001} W. Souma, Y. Fujiwara, and A. Aoyama, {\it cond-mat/0108482}.
\bibitem{Galaschelli2008} D. Garlaschelli and M. I. Loffredo, J. Phys. A
	{\bf 41} 224018(2008).
\bibitem{Gnedenko} B. V. Gnedenko and A. N. Kolmogorov, {\it Limit
	distributions for sums of independent random variables},
	Addison-Wesley, 1954.
\bibitem{Barabasi1999}  A.-L. Barab\'asi and R. Albert,  Science {\bf
	286} 509(1999).
\bibitem{Chebyshev} P. L. Chebyshev, Acta. Math. {\bf 14} 305(1890).
\end{thebibliography}
\end{document}